\documentclass[
12pt]{article}
\pdfoutput=1
\usepackage{epsfig}
\usepackage{graphicx}

\makeatletter

\@addtoreset{equation}{section}
\makeatother
\newcommand{\be}{\begin{equation}}
\newcommand{\ee}{\end{equation}}
\newcommand{\ben}{\begin{equation*}}
\newcommand{\een}{\end{equation*}}
\newcommand{\bea}{\begin{eqnarray}}
\newcommand{\eea}{\end{eqnarray}}
\newcommand{\alg}{\begin{align}}
\newcommand{\algx}{\end{align}}

\begin{document}

\begin{titlepage}
\vskip1cm
\begin{flushright}
\end{flushright}
\vskip1.25cm
\centerline{\large
\bf  Information metric and Euclidean Janus correspondence}
\vskip1.27cm \centerline{ \large Dongsu Bak }
\vspace{1.25cm} \centerline{\sl  Physics Department,
University of Seoul, Seoul 02504 Korea}
 \vskip0.3cm
 \centerline{\sl
Fields, Gravity \& Strings, Center for Theoretical Physics of the Universe}
 \centerline{\sl Institute for Basic Sciences, Daejeon 34047 Korea}
\vskip0.3cm
 \centerline{\tt ($\,$dsbak@uos.ac.kr$\,$)} \vspace{1.5cm}

\centerline{ABSTRACT}
 \vspace{0.75cm} \noindent
We consider the quantum information metric of a family of CFTs perturbed by an exactly marginal operator,  
which has the dual description of   the Euclidean Janus geometries. 
We first clarify  its two dimensional case dual to the three dimensional Janus geometry, which recently appeared in arXiv:1507.07555. 
We generalize this correspondence to higher dimensions and get a 
precise agreement between the both sides. We also show that the mixed-state information metric of 
the same family of CFTs has a dual description in the Euclidean version of the Janus time-dependent 
black hole geometry.

\vspace{1.75cm}
\end{titlepage}

\section{Introduction}

There has been a lot of progress in understanding the AdS/CFT correspondence  \cite{Maldacena:1997re}. 
However  the details of the dictionary in application of decoding general gravity backgrounds are far from complete. 
The recent proposal   \cite{MIyaji:2015mia} of quantum information metrics and their dual gravity descriptions  (see also \cite{kk} for the related works)
is one such example of new dictionary which may shed new light on this issue.  

Information metric or fidelity susceptibility of quantum system measures the distance between two infinitesimally  different quantum states
$| \psi(\lambda )\rangle$ and $|\psi(\lambda+\delta \lambda)\rangle$ where $\lambda$ is a parameter 
describing a family of perturbed quantum system (see \cite{caves} and the review \cite{Gu}). The fidelity of the 
perturbation is defined by the inner product of these states, 
\be
{\cal F} (\lambda,\lambda+\delta \lambda)
\equiv  |\langle \psi(\lambda)| \psi(\lambda+\delta\lambda )\rangle | = 1- G_{\lambda\lambda} \delta \lambda^2
+O(\delta \lambda^3)
\ee
measuring the overlap of the two states where  the quantum information metric $G_{\lambda \lambda}$ is defined as 
the  coefficient  of 
$-\delta\lambda^2$.
This has an application
in understanding the quantum phase transitions or, in general, characteristic response of quantum system under some parametric 
perturbation. One of the prime examples of this correspondence in  \cite{MIyaji:2015mia} involves two dimensional CFTs perturbed by an exactly 
marginal scalar operator and their gravity dual given by the Euclidean three dimensional  Janus system. Originally the Janus
 geometry  \cite{Bak:2003jk} is dual to an
interface CFT  where the boundary values of the scalar field dual to an exactly marginal operator jump across the interface and has
many other applications including the physics of  Janus time-dependent black holes (TDBH) \cite{Bak:2007jm,Bak:2007qw}.

In this note, we would like to extend  this correspondence in two ways. First, we generalize the above correspondence to arbitrary dimensions. Namely the information metric
of $d+1$ dimensional CFTs perturbed by an exactly marginal scalar operator will be dual to the $d+2$ dimensional  Euclidean Janus geometry. 
Below we shall make this clear by matching the two sides in a rather precise manner.
Next we consider the mixed-state information metric \cite{Uhlmann}
 that measures
an infinitesimal distance between two thermal states  labeled by perturbations of the underlying   quantum system\footnote{See also \cite{Alishahiha:2015rta} for the recent discussion of the mixed-state information metric in relation with holography.}. In the field theory side,
we shall  consider again two dimensional CFTs perturbed by an exactly marginal scalar operator but, now, at finite temperature.
 The thermal state 
will be described by the thermal density matrix whose trace corresponds to the standard thermal partition function. 
Its gravity dual will be identified with the Euclidean Janus time-dependent black hole. Indeed the Lorentzian Janus TDBH involves two 
causally separate 
boundary spacetimes, on each of which a CFT lives \cite{Bak:2007qw}.  The values of the coupling of the exactly marginal 
scalar operator of these CFTs  differ from each other. In the field theory side, the system is described by the tensor product state of the two CFTs which is 
entangled initially in a manner appropriate to describe the mixed state fidelity. From the view point of the one boundary CFT, its time evolution describes the thermalization of an initial 
out-of-equilibrium perturbation \cite{Bak:2007qw}.  
We shall argue that the Euclidean version of this Janus TDBH geometry describes the mixed-state information metric if one introduces a regularization
in a particular manner. We further generalize this Euclidean Janus TDBH correspondence to higher dimensions where the corresponding CFTs are 
defined on $S^1 \times \Sigma_d$ with $d >1$ where $\Sigma_d$ is the hyperbolic space which can be made to be compact by an 
appropriate quotient
by its symmetry action.

\section{Information metric and Euclidean Janus}
We begin by first clarifying the analysis in Ref.~\cite{MIyaji:2015mia} for the 2D case which  has the dual description of the 3D Euclidean Janus 
geometry. We then 
generalize this correspondence to higher dimensions. Let $|\Omega_i \rangle \, (i=1,2)$ describe
the ground state of $d+1$ dimensional CFT$_i$ on $R^{1+d}$ whose Lagrangian density and Hamiltonian are denoted by
${\cal L}_i(\lambda_i)$ and $H_i(\lambda_i)$ respectively. Then the fidelity is defined by 
\be
{\cal F}=|\langle \Omega_2 | \Omega_1 \rangle |= \frac{1}{(Z_1 Z_2)^{\frac{1}{2}}} \int {\cal D} 
\Phi \, e^{- \int d^d x \left[ \int^\infty_{0} d\tau {\cal L}_2 + \int^0_{-\infty} d\tau {\cal L}_1 
\right]}
\ee 
where $\Phi$ collectively denotes the field content of the underlying field-theory and $Z_i$ is the partition function of the corresponding 
Euclidean system. Let us assume the perturbation
is given by the primary operator as
\be
{\cal L}_2 -{\cal L}_1 = \delta \lambda \,  O(\tau, x)
\ee
with $\delta\lambda = \lambda_2 -\lambda_1$ and we would like to compute the regularized inner product ${\cal F}_{\epsilon}= |\langle \Omega_2 (\epsilon) | \Omega_1 \rangle | $ where
the regularized state $ | \Omega_2 (\epsilon) \rangle $ is defined by
\be
 | \Omega_2 (\epsilon) \rangle  =\frac{ e^{-\epsilon H_1} | \Omega_2  \rangle}{(\langle \Omega_2 | e^{-2 \epsilon H_1} |\Omega_2 \rangle)^{\frac{1}{2}}}
\ee
With this set-up, the information metric is identified in terms of  the two-point correlation function as
\be
G_{\lambda\lambda} = \frac{1}{2}   \int_\epsilon^\infty d \tau_2  \int^{-\epsilon}_{-\infty} d\tau_1  \int d^d x_2 \int d^d x_1
\langle  O(\tau_2, x_2) O(\tau_1, x_1)  \rangle 
\ee
We then specialize to the case of scalar primary operator of dimension $\Delta$ for which the two-point function is given by
\be
\langle O(\tau, x) O(\tau', x')\rangle  = \frac{{\cal N}_{\Delta}} {\left[ \, (\tau-\tau')^2 +(x-x')^2 \,  \right]^\Delta}
\ee 
In order to match  with the gravity description, we shall take the normalization  
\be
{\cal N}_\Delta= \frac{2 \eta\ \ell^{d}\, (d+1) \, \Gamma(\Delta ) }{\pi^{\frac{d +1}{2}} \Gamma(\Delta -\frac{d+1}{2})  }
\label{normalization}
\ee
where $\eta= \frac{1}{16\pi G}$, with the $d+2$ dimensional Newton's constant $G$, and $\ell$ is the AdS radius scale. In the AdS/CFT 
correspondence, there exists a bulk scalar field dual to the scalar primary operator $O$ of dimension $\Delta$ 
and the above normalization follows from the bulk scalar action whose kinetic term is normalized as
$I_\phi= +\eta \int d^{d+2} x \sqrt{g} (\nabla \phi)^2 + \cdots$ \cite{Bak:2017rpp}. By the straightforward computation,
the information metric of the perturbation can be identified with \footnote{The result here agrees with that of Ref.~\cite{MIyaji:2015mia} 
after taking care of the 
normalization of the two-point correlation function in (\ref{normalization}).}
\bea
G_{\lambda\lambda}= \frac{\eta \ell^d \, V_d}{2\sqrt{\pi}}\, \frac{ (d+1)\,  \Gamma(\Delta -\frac{d}{2}-1)}
{(2\Delta -d-1) \Gamma(\Delta -\frac{d+1}{2}) }\, \frac{1}{(2\epsilon)^{2\Delta -(d+2)}}
\eea
where $V_d$ is the spatial volume of $R^d$. We assumed here $2\Delta -(d+2) >0$, otherwise one needs further 
IR cut-off dependence.    
When the deformation  is by an  exactly marginal operator of dimension $\Delta=d+1$, the expression  is further reduced to
\be
G_{\lambda\lambda}= \frac{\eta \ell^d \, V_d}{2\sqrt{\pi}}\, \frac{\Gamma(\frac{d}{2})}
{ \Gamma(\frac{d+1}{2}) }\, \frac{1}{(2\epsilon)^{d}}
\label{exact}
\ee
which will be compared to the gravity side  below. In these expressions, we see that the information metric scales linearly as the
 spatial volume $V_d$
of the spacetime on which the state is defined. 

Let us now turn to the dual gravity side. The dual geometry is described by the AdS Einstein-scalar system described by the action 
\be
I= -\frac{1}{16\pi G}\int d^{d+2} x \sqrt{g} \left[ R -g^{ab} \partial_a \phi \partial_b \phi + \frac{d (d+1)}{\ell^2} \right]
\label{einstein}
\ee
where $\ell$ is the AdS radius scale. 
In three and five dimensions, this system can be consistently embedded into the type IIB supergravity 
and, hence, via the standard  AdS/CFT correspondence, the microscopic understanding of the underlying 
system is allowed \cite{Bak:2003jk, Bak:2007jm}. But we shall not
discuss this direction any further in this note. The scalar field  here corresponds to an exactly marginal scalar operator in the field-theory side, 
which is further identified as a Lagrange density operator in the above type  IIB supergravity examples.   

For three dimensional case, we review the computation in Ref.~\cite{MIyaji:2015mia} with a slight refinement.
The relevant  Janus solution is given by
\bea
&& ds^2=\ell^2 \left[ dy^2 + f(y) \, ds^2_{AdS_2}\right]                        \cr
&& \phi(y)= \phi_0+ \frac{1}{\sqrt{2}} \log\left( 
\frac{1+\sqrt{1-2\gamma^2} +\sqrt{2} \gamma \tanh y}
{1+\sqrt{1-2\gamma^2} -\sqrt{2} \gamma \tanh y }
\right)
\label{3djanus}
\eea
where $f(y)= \frac{1}{2}(1+\sqrt{1-2\gamma^2} \cosh 2y )$ with $\gamma < \frac{1}{\sqrt{2}}$ 
and we choose the AdS$_2$ metric that of the Euclidean Poincar\'e patch given by 
$ds_{AdS_2}= \frac{d\xi^2 +dx^2}{\xi^2}$. Since $\phi(\pm \infty)=\phi_0 \pm  \frac{1}{\sqrt{2}} \tanh^{-1} \sqrt{2} \gamma$, 
one finds
\be
\delta \lambda=  \lambda_2- \lambda_1 = 2 \gamma +O(\gamma^3) 
\label{delta}
\ee 
with the identification $\phi(-\infty)= \lambda_1$ and $\phi(\infty)= \lambda_2$.
For the regularization of the on-shell gravity action, we introduce a pure AdS metric given by
\be
 d\hat{s}^2=\ell^2 \left[ d\hat{y}^2 +\frac{1}{2}(1+ \cosh 2\hat{y} ) \, ds^2_{AdS_2}\right]    
\ee 
To match the Janus geometry at $\pm y_\infty$ with the above reference metric at $\pm \hat{y}_\infty$, 
the matching condition can be read off as
\be
\sqrt{1-2\gamma^2} \, \cosh 2y_\infty =  \cosh 2\hat{y}_\infty
\ee 
With this  preparation, the on-shell gravity action is evaluated as
\bea
I_\gamma = \frac{\ell}{4\pi G} V_{AdS_2} \Gamma_\gamma
\eea
where the factor $\Gamma_\gamma $ is given by  the integral 
\be
\Gamma_\gamma=\int^{y_\infty}_{-y_\infty} dy  \frac{1}{2}(1+\sqrt{1-2\gamma^2} \cosh 2y )=
 y_\infty +\frac{1}{2} \sqrt{1-2\gamma^2} \sinh 2y_\infty 
\ee
We shall subtract the reference  contribution of $\Gamma_0= \hat{y}_\infty +\frac{1}{2} \sinh 2 \hat{y}_\infty $. The AdS volume factor
will  also be regularized as
\be
V_{AdS_{2}}= \int dx \int^\infty_{\epsilon_0} \frac{d\xi}{\xi^2} =\frac{V_1}{\epsilon_0}
\ee
where $\epsilon_0$ is the cut-off in the gravity side. Thus  the difference of the on-shell gravity action is evaluated as
\be
I_\gamma -I_0 = \frac{\ell}{16\pi G}\,  \frac{V_1}{\epsilon_0}\, \log\frac{1}{1-2\gamma^2}
\ee 
Using ${\cal F}^{bulk}_{\epsilon_0}= e^{-(I_\gamma -I_0)}$, the corresponding information metric is identified as
\footnote{This does not agree with the result in Ref.~\cite{MIyaji:2015mia}, where we believe the authors  missed the factor of $2$ in (\ref{delta}).}
\be
{G}^{bulk}_{ \lambda \lambda}= \frac{\eta \ell}{2}\, V_1 \, \frac{1}{\epsilon_0}
\ee
which agrees with the field-theory computation in (\ref{exact}) if one identifies $\epsilon_0 = 2 \epsilon$. This identification of the cut-off scales may look 
rather ad hoc since  any numerical discrepancy may be absorbed into the proportionality coefficient. However, we shall show 
below  that precise match holds even for the 
general dimensional case with the same identification of $\epsilon$ and $\epsilon_0$. 
One comment is that the above can be given in terms of purely field-theoretic quantity
noting the relation $\eta \ell  = \frac{c}{24 \pi}$ where $c$ is the central charge of the two dimensional CFT. 

For the general dimensions, again the gravity dual of an exactly marginal operator deformation is 
described by the gravity action in (\ref{einstein}). The relevant
Euclidean Janus solution can be given as
\bea
ds_\pm^2\ &=& \frac{\ell^2}{q^2_\pm} \left[\frac{ dq^2_\pm}{P(q_\pm)} +   ds^2_{AdS_{d+1}}\right]     \cr
\phi(q_\pm)&=& \phi_0  \pm \gamma \int_{q_\pm}^{q_*} dg  \frac{ g^{d}}{\sqrt{P(g)}}
\label{d+2janus}
\eea 
where $P(g)=1-g^2 + \frac{\gamma^2}{d(d+1)} g^{2d+2}$ and $q_*$ denotes the smallest positive root of $P(q_*)=0$. $q_\pm$ is 
ranged over $[0, q_*]$ and
$\gamma$ here is the 
deformation parameter, which is ranged over the interval $[\, 0,\sqrt{d}\left(\frac{d}{d+1}\right)^{\frac{d}{2}}\,)$.  We shall again choose 
the AdS$_{d+1}$ part as the Euclidean Poincare patch 
whose metric is 
given by $ds^2_{AdS_{d+1}}= \frac{d\xi^2 + d \vec{x}_d \cdot d \vec{x}_d}{\xi^2}$. In this geometry, $g_\pm\rightarrow 0$ correspond  to 
asymptotic regions where the values of the scalar field become $\phi_+ =  \lambda_2$ and $\phi_- = \lambda_1$, respectively describing the deformations of
the boundary CFT by the corresponding operator. We shall match the $+$ and the $-$ patches at $q_\pm =q_*$ where the geometry can be smoothly joined.
There are many ways of parameterizing the Janus solutions but, for the present purpose, we find that  the above explicit one is most convenient. 
To the leading orders of $\gamma$, the difference of the boundary values of the scalar field is given by
\be
\delta  \lambda=  \lambda_2- \lambda_1 = \gamma \frac{\sqrt{\pi} \, \Gamma(\frac{d+1}{2})}{\Gamma(\frac{d}{2}+1)}+O(\gamma^3)
\label{diff}
\ee 
As in the $d=1$ case, the on-shell action can be evaluated as
\be
I_\gamma = 2(d+1)\,  \eta \ell^{d} \, V_{AdS_{d+1}}\, \Gamma_\gamma
\ee
with 
\be
\Gamma_\gamma= 2 \int^{q_*}_{\epsilon_q} \frac{dg}{ g^{d+2} \sqrt{P(g)}}
\ee
 where we introduce a cut-off $q^{as}_{\pm}= \epsilon_q$
to regulate the infinite contribution of the asymptotic region. The way to proceed is the same as before. We introduce the reference metric
of pure AdS spacetime
\be
d\hat{s}_\pm^2\ = \frac{\ell^2}{\hat{q}^2_\pm} \left[\frac{ d\hat{q}^2_\pm}{1-\hat{q}^2_\pm} +   ds^2_{AdS_{d+1}}\right] 
\ee 
and match the Janus geometry at $q^{as}_\pm$ with the reference at $\hat{q}^{as}_{\pm}$ 
where the matching condition reads $q_\pm^{as}=\hat{q}^{as}_{\pm}=
\epsilon_q$.  For the power series expansion of $\Gamma_\gamma $ in terms of $\gamma$, we find it is convenient to introduce a new 
integration variable $\chi= \frac{g^2}{q^2_*}$. Then $\Gamma_\gamma$ can be rewritten as
\be
\Gamma_\gamma= \frac{1}{q^{d+1}_*}\int^{1}_{\frac{\epsilon^2_q}{q^2_*}} \frac{d\chi \, \chi^{-\frac{d+3}{2}}}{ 
\sqrt{ (1-\chi) \left( 1-\frac{\gamma^2 \, q_*^{2d+2}}{d(d+1)} \frac{\chi^{d+1}-\chi}{\chi-1}  \right) }}
\ee
Further noting $q^2_* = 1+ \frac{\gamma^2}{d(d+1)} +O(\gamma^4)$, it is straightforward to expand $\Gamma_\gamma$ as a power series of
$\gamma$. With some further algebra, the difference becomes
\be
\Gamma_\gamma - \Gamma_0 = \frac{\gamma^2}{d(d+1)}\frac{
 \,_2F_1(-\frac{1}{2}, \frac{1-d}{2},\frac{1}{2},1-\epsilon_q^2 )}{\sqrt{1-\epsilon_q^2} }= \frac{\gamma^2}{d(d+1)}  
\frac{\sqrt{\pi} \, \Gamma(\frac{d+1}{2})}{\Gamma(\frac{d}{2})}+O(\epsilon_q^2)
\ee
to the order $\gamma^2$ where $\,_2F_1(a,b,c,x)$ is the hypergeometric function.
Since the leading singular contributions of $\Gamma_0$ and $\Gamma_\gamma$ are of order $\epsilon^{-(d+1)}_q$, there are many 
additional singular terms in  $\Gamma_0$ and $\Gamma_\gamma$ in general. Then the above computation implies that, up to the 
order $\gamma^2$,
all those coefficients of the singular terms in    $\Gamma_0$ and $\Gamma_\gamma$ cancel against with one another precisely . 
We find this fact remarkable! We regulate the AdS volume in the same way as before;  The resulting volume is evaluated as
\be
V_{AdS_{d+1}}= V_d \int^\infty_{\epsilon_0} \frac{d\xi}{\xi^{d+1}}= \frac{V_d}{d\, \epsilon^d_0} 
\ee
From ${\cal F}^{bulk}_{\epsilon_0}= e^{-(I_\gamma -I_0)}$ together with  the relation (\ref{diff}), the information metric is identified as
\be 
G^{bulk}_{ \lambda \lambda} = \frac{\eta \ell^d}{2}\,   \frac{ \Gamma(\frac{d}{2})}{\sqrt{\pi}\, \Gamma(\frac{d+1}{2})}  \, \frac{V_d}{\epsilon^d_0}
\ee  
which agrees with the field-theory result in (\ref{exact}) if one identifies $\epsilon_0 =2\epsilon$.

\section{Janus TDBH and mixed-state information metric}

The mixed-state fidelity is similarly defined by 
\be
{\cal F}(\lambda_1,\lambda_2) = {\rm tr} \, \left[ \rho_1^{\frac{1}{4}}\rho_2^{\frac{1}{2}} \rho_1^{\frac{1}{4}} \right]
\ee
where $\rho_i$  represents a thermal density matrix given by $\rho_i = \frac{e^{-\beta_i H_i(\lambda_i)}}{Z_i}$
and $\lambda$ is parameterizing  perturbation by some operator of the underlying  quantum system. 
The temperatures may differ from each other in general but we shall 
specialize in the case where $\beta_1= \beta_2 = \beta$. The mixed state information metric is defined by \cite{Uhlmann}
\be
{\cal G}_{\lambda \lambda} = -\lim_{\delta \lambda \rightarrow 0} \frac{\log {\cal F}(\lambda, \lambda+ \delta \lambda)}
{\delta\lambda^2}
\ee
which measures an infinitesimal distance between the two thermal states. We consider again the case where the underlying theory is given by the 
CFT which is perturbed by some primary operator with ${\cal L}_2 -{\cal L}_1= \delta \lambda \,  O(\tau,x)$ as before. We shall 
introduce a regularization of the second density matrix by
\be
\rho^{\frac{1}{2}}_2 (\epsilon) = \frac{e^{-\epsilon H_1 }e^{-\left(\frac{\beta}{2}-2\epsilon\right) H_2} e^{-\epsilon H_1 }}{\sqrt{Z_2(\epsilon)}}
\ee
where $Z_2(\epsilon)={\rm tr}\,  
(e^{-\epsilon H_1 }e^{-\left(\frac{\beta}{2}-2\epsilon\right) H_2} e^{-\epsilon H_1 })^2 $. Compared to the pure state case, the cut-off looks
a bit unusual since its total size is doubled. This may be understood as follows; One can view that $\rho_1^{\frac{1}{2}}$ is defined over the 
interval ${\cal I}_1=[-\frac{\beta}{2}, 0]$ and  $\rho_2^{\frac{1}{2}}$   over  ${\cal I}_2 =[0, \frac{\beta}{2}]$ 
covering full thermal-circle of $S^1$ which is 
$\beta$ periodic. To be consistent with the periodicity, one has to introduce the cut-off at both ends of the intervals ${\cal I}_1$ and ${\cal I}_2$ specified respectively by ${\cal I}_1(\epsilon) =[ -\frac{\beta}{2}+\epsilon, -\epsilon]$ and 
${\cal I}_2(\epsilon) =[\epsilon, \frac{\beta}{2}-\epsilon]$ since $\frac{\beta}{2}\sim -\frac{\beta}{2}$ due to the periodicity.
With this preliminary, the mixed-state information metric is identified as
 \be
{\cal G}_{ \lambda \lambda} = \frac{1}{2} \int_\epsilon^{\frac{\beta}{2}-\epsilon} d \tau_2 
\int^{-\epsilon}_{-\frac{\beta}{2}+\epsilon} d\tau_1 \int d^d x_2 \int d^d x_1 
\langle \hat{O}(\tau_2, x_2) \hat{O}(\tau_1, x_1)\rangle
\ee
with $ \hat{O}(\tau, x) \equiv  O(\tau, x)-  \langle O(\tau, x) \rangle $. Here the average is defined with the thermal density matrix
$\rho_1$ 
as $\langle\  \bullet  \ \rangle= {\rm tr}   \, \bullet \,   \rho_1$.
To evaluate the above mixed-state 
information metric, we  assume that the one-point function vanishes and specialize in the case of the two dimensional  ($d=1$) CFT.   
The two-point function is given by
\be
\langle O(\tau, x) O(\tau', x')\rangle  = \frac{{\cal N}_{\Delta} \left(\frac{\sqrt{2} \pi}{ \beta}\right)^{2\Delta}} 
{\left[ \cosh \left( \frac{2\pi}{\beta}(x-x')\right)
- \cos \left( \frac{2\pi}{\beta} (\tau-\tau')\right)  \,  \right]^{\Delta}}
\ee
which appeared in \cite{KeskiVakkuri:1998nw, Maldacena:2001kr} in the AdS/CFT context. Then the corresponding information metric can be 
evaluated in a straightforward manner. Again for $2\Delta -(d+2) >0$ with $d=1$, its most singular part comes from the regions 
near $x_1-x_2 \sim 0, \ \tau_1-\tau_2 \sim 0$ or $\beta$. Therefore, 
one has ${\cal G}_{\lambda \lambda}(d=1)=2G_{\lambda \lambda}(d=1)+\cdots$  
 where dots represent the less singular contributions.
 
 For general dimensions with $d >1$, one can argue that the short-distance structure of the two-point function remains the same 
 as that of the pure state case. Since the short-distance regions of the thermal two-point functions occur near 
 $\vec{x}_1-\vec{x}_2 \sim 0, \ \tau_1-\tau_2 \sim 0$ or $\beta$, one finds again ${\cal G}_{\lambda \lambda}(d)=
 2G_{\lambda \lambda}(d)+\cdots$ even for the higher dimensions. 
 
 In the gravity side, we shall focus on the case of CFT system perturbed by an exactly marginal operator 
 as before.  Below we shall argue that the natural candidate of the gravity dual of the mixed-state information
 is given by the Euclidean version of the Janus TDBH solution, which is the solution of the Einstein scalar system
 in (\ref{einstein}).  It still takes the form in (\ref{3djanus})
\be
ds^2= dy^2 + f(y) ds^2_{M_2}
\ee
 but the $AdS_2$ part is replaced by the two metric
 \be
 ds^2_{M_2} = d \rho^2 +(\cosh \rho)^2  \, \alpha^2 dx^2 
 \ee
where $-\infty <\rho < \infty$. The form of the scalar field $\phi(y)$ also remains the same as that in (\ref{3djanus}). 
$\alpha$ here may be taken arbitrary but is chosen as $\alpha=\frac{2\pi}{\beta}$ in order 
to match the coordinate $x$ with that of the boundary CFT. For general dimensions with $d>1$,  while all the rest intact, we simply replace 
the AdS$_{d+1}$ part in (\ref{d+2janus}) by
\be
ds^2_{M_{d+1}}=  d \rho^2 +\cosh^2 \rho  \, ds_{\Sigma_d}^2 
\ee
 where $ ds_{\Sigma_d}^2 $ is the $d$ dimensional hyperbolic space, with unit radius, which can be made compact by an appropriate 
quotient. Then this higher dimensional Euclidean Janus TDBH solution can be dealt by the treatment below but let us here focus on the
$d=1$ case for definiteness.

\begin{figure}[ht!]
\centering  
\includegraphics[width=5cm]{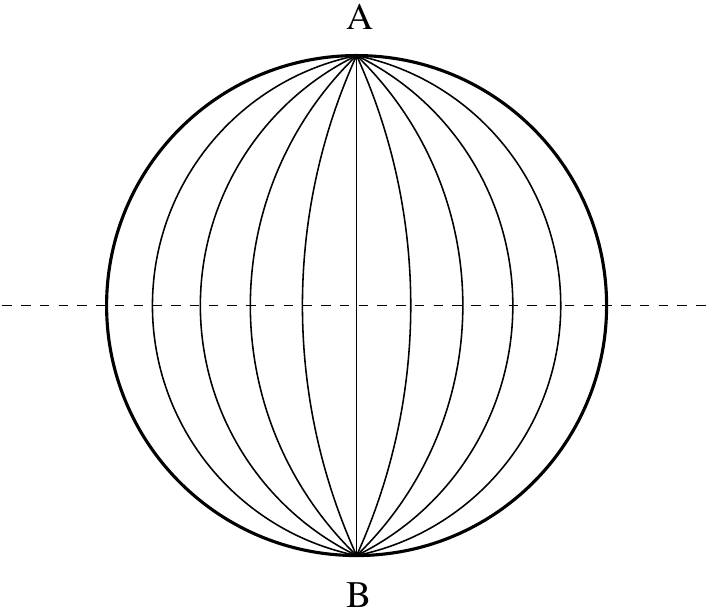}
\caption{\small The conformal shape of the Euclidean time-dependent black hole in ($y,\  \nu $) space is depicted. $y=-\infty$  corresponds to 
the left semicircle boundary from $B$ and $A$ while  $y=\infty$  to the right semi-circle boundary also from $B$ to $A$. The Euclidean time coordinate 
$\tau$  
runs from $B$ to $A$. 
}
\label{fig3}
\end{figure}

  Introducing a new coordinate $\nu$ defined by $\cosh \rho = \frac{1}{\sin \nu}$, the above two metric 
becomes
\be
 ds^2_{M_2} = \frac{d \nu^2   + \alpha^2 dx^2 }{\sin^2 \nu}
\ee 
where $0 \le \nu \le \pi$. In Figure 1, we depict the conformal shape of the Euclidean geometry in ($y, \nu$) space. 
There the $\nu$ coordinate  runs upward from the point $B$ to $A$ whereas the $y$ 
coordinate from the left  to the right semi-circles. Namely $y= -\infty$ corresponds to the left boundary whereas the $y=\infty$ 
corresponds to the right boundary, which are joined at the points $A$ and $B$.  We introduce the Euclidean time coordinate 
$\tau=\frac{\nu}{\alpha} \in [\epsilon_0,\frac{\beta}{2}-\epsilon_0 ]$ where 
$\epsilon_0$ is the cut-off in the gravity side.  In relation with the boundary field-theories,  the  left boundary Euclidean time coordinate is given 
by $\tau_1=-\tau \in  [ -\frac{\beta}{2}+\epsilon_0, -\epsilon_0 ]$ 
and the right by $\tau_2=\tau \in [\epsilon_0,\frac{\beta}{2}-\epsilon_0 ]$ with the cut-off introduced. The evaluation of the Euclidean 
action can be performed in a similar manner as before leading to
\be
I_\gamma = \frac{\ell}{4\pi G} V_{M_2} \Gamma_\gamma
\ee 
The regulated volume reads
\be
V_{M_2} = V_1 \alpha^2  \int^{\frac{\beta}{2}-\epsilon_0}_{\epsilon_0} \frac{d\tau}{\sin^2 \alpha \tau} = \frac{2 V_1}{\epsilon_0}+\cdots
\ee
where dots represent any nonsingular contributions. Therefore, again from  ${\cal F}^{bulk}_{\epsilon_0}= e^{-(I_\gamma -I_0)}$, one finds
${\cal G}^{bulk}_{\lambda \lambda}= 2 G^{bulk}_{{\lambda \lambda}}+\cdots$, which agrees with the CFT result with the identification $\epsilon_0 =2\epsilon$.
The similar analysis goes through for the higher dimensional case where $V_d$ should be interpreted as the volume of the hyperbolic 
space divided by $\alpha^d$ but we shall not go to this direction any further.
For the match of the finite parts or in general the less singular contributions of the gravity and the CFT results, perhaps one 
needs a better way of matching the regularization of the both sides. The regularization here does not appear to be related to any
renormalization of the underlying system. We leave further investigation of this issue for the future problem.

The Lorentzian geometry is given by the analytic continuation $\rho\rightarrow i p$ and the metric takes the form 
\be
ds^2_{M_{1,1}} = - dp^2 + (\cos p)^2 \, \, \alpha^2 d x^2 
\ee
where $ -\frac{\pi}{2} \le p \le \frac{\pi}{2}$. The boundary time coordinate is related by $\sin p  = \tanh \alpha t $ 
with $ t \in (-\infty, \infty)$.  This Lorentzian geometry has a time reflection $Z_2$ symmetry of $t \rightarrow -t$. The lower half of the
Euclidean geometry in Figure 1 can be smoothly joined to the upper half of the Lorentzian geometry with $t \ge 0$. This picture provides
us with the initial quantum state dual to the geometry \cite{Bak:2007qw} following the idea of the Hartle-Hawking construction of the 
wavefunction \cite{Hartle:1983ai}. Since there are two separate boundary spacetimes in the Lorentzian geometry, the relevant Hilbert space 
consists of ${\cal H}_1 \times {\cal H}_2$. Using the standard dictionary of the AdS/CFT correspondence of the Euclidean geometry,
the Euclidean time evolution along the boundary of the lower half geometry is identified as the evolution operator
$U= e^{-\frac{\beta}{4} H_2} e^{-\frac{\beta}{4} H_1}$. Thus the initial state of the Lorentzian geometry at $t=0$ becomes
\be
|\psi \rangle = \frac{1}{\sqrt{Z}}\sum_{mn} \langle E_m^{(2)} | U | E_n^{(1)}\rangle\, | E_n^{(1)}\rangle \times | E_m^{(2)}\rangle
\ee
where $|E_n^{(i)}\rangle $ is the energy eigenstate of $H_i$ respectively and 
$Z= {\rm tr}\, U^\dagger U$ (see \cite{Maldacena:2001kr} for  the construction initial state for the case of BTZ black hole). Further Lorentzian time evolution is given by
$|\psi (t_1, t_2)\rangle = e^{-i (t_1 H_1 \times I + t_2 I \times H_2 )} |\psi \rangle $ where $t_i$ is the $i$-th boundary time 
coordinate. This time-dependent geometry describes the physic of thermalization. The geometry in the far future approaches that of the BTZ black hole with the equilibrium temperature $\beta^{-1}$. The one-point function of the exactly marginal operator dual to the scalar field can be computed from
the both sides leading to \cite{Bak:2007qw} 
\be
\langle O_{\Delta=2}(it,0)\rangle = \frac{ c}{12 \pi } \, \frac{\gamma\, \alpha^2}{\cosh^2 \alpha t}   
\ee  
Thus the geometry describes the thermalization of an initially out-of-equilibrium state by an exponential relaxation
of the perturbation in the far future.

From this construction, it is clear that $Z=e^{-I}$ can be identified as the partition function of the full Euclidean geometry including the lower and the 
upper halves at the same time. Then the mixed-state fidelity that is defined by $\frac{Z}{\sqrt{Z_1 Z_2}}$ is naturally identified as
\be
{\cal F}^{bulk} = e^{-\left({I-\frac{I_1+I_2}{2}}\right)}
\ee
in the gravity side where the relation $Z_i =e^{-I_i}$ is used. Hence the problem is reduced to the regularization
 problem to match the less singular contributions of the both sides,
which presumably  requires a set-up different from that of the present note. Further investigation is required in this direction. 

\section*{Acknowledgement}
We would like to thank Choonkyu Lee for the valuable comments and  careful reading of the manuscript.
This work was
supported in part by NRF Grant 2014R1A1A2053737.


\begin{thebibliography}{99}

\bibitem{Maldacena:1997re}
  J.~M.~Maldacena,
  ``The large N limit of superconformal field theories and supergravity,''
  Adv.\ Theor.\ Math.\ Phys.\  {\bf 2}, 231 (1998)
  [Int.\ J.\ Theor.\ Phys.\  {\bf 38}, 1113 (1999)]
  [arXiv:hep-th/9711200].
  
\bibitem{MIyaji:2015mia} 
  M.~Miyaji, T.~Numasawa, N.~Shiba, T.~Takayanagi and K.~Watanabe,
 ``Gravity Dual of Quantum Information Metric,''
  arXiv:1507.07555 [hep-th].

\bibitem{kk} 
  M.~Nozaki, S.~Ryu and T.~Takayanagi,
  ``Holographic Geometry of Entanglement Renormalization in Quantum Field Theories,'' 
 JHEP {\bf 1210}, 193 (2012)  
 [arXiv:1208.3469 [hep-th]];
  M.~Miyaji and T.~Takayanagi,
  ``Surface/State Correspondence as a Generalized Holography,''  
PTEP {\bf 2015}, no. 7, 073B03 (2015)  
[arXiv:1503.03542 [hep-th]];
  M.~Miyaji, T.~Numasawa, N.~Shiba, T.~Takayanagi and K.~Watanabe,
  ``Continuous Multiscale Entanglement Renormalization Ansatz as Holographic Surface-State Correspondence,''  
Phys.\ Rev.\ Lett.\  {\bf 115}, no. 17, 171602 (2015) 
[arXiv:1506.01353 [hep-th]].  



\bibitem{caves} S. L. Braunstein and C. M. Caves, 
``Statisical distance and the geometry of quantum states", 
Phys. Rev. Lett. 
{\bf 72} (1994) 3439.

\bibitem{Gu}
S-J Gu, 
``Fidelity approach to quantum phase transition", 
Int. J. Mod. Phys. B {\bf 24} (2010) 4371.





\bibitem{Bak:2003jk} 
  D.~Bak, M.~Gutperle and S.~Hirano,
  ``A Dilatonic deformation of AdS(5) and its field theory dual,''
  JHEP {\bf 0305}, 072 (2003)
  [hep-th/0304129].

\bibitem{Bak:2007jm}
  D.~Bak, M.~Gutperle and S.~Hirano,
  ``Three dimensional Janus and time-dependent black holes,''
  JHEP {\bf 0702} (2007) 068
  [arXiv:hep-th/0701108].

\bibitem{Bak:2007qw}
  D.~Bak, M.~Gutperle and A.~Karch,
  ``Time dependent black holes and thermal equilibration,''
  JHEP {\bf 0712}, 034 (2007)
  [arXiv:0708.3691 [hep-th]].


\bibitem{Uhlmann} A. Uhlmann, 
"The transition probability in the state space of a *-algebra"
Rep. Math. Phys. {\bf 9} 273 (1976). 


\bibitem{Alishahiha:2015rta} 
  M.~Alishahiha,
 ``Holographic Complexity,''
  Phys.\ Rev.\ D {\bf 92}, no. 12, 126009 (2015)
  [arXiv:1509.06614 [hep-th]].


\bibitem{Bak:2017rpp} 
  D.~Bak and A.~Trivella,
  ``Quantum Information Metric on R X S(d-1),''
  arXiv:1707.05366 [hep-th].





\bibitem{KeskiVakkuri:1998nw} 
  E.~Keski-Vakkuri,
  ``Bulk and boundary dynamics in BTZ black holes,''
  Phys.\ Rev.\ D {\bf 59}, 104001 (1999)
  [hep-th/9808037].


\bibitem{Maldacena:2001kr}
  J.~M.~Maldacena,
  ``Eternal black holes in anti-de Sitter,''  
  JHEP {\bf 0304}, 021 (2003)  [hep-th/0106112].  



\bibitem{Hartle:1983ai} 
  J.~B.~Hartle and S.~W.~Hawking,
  ``Wave Function of the Universe,''
  Phys.\ Rev.\ D {\bf 28}, 2960 (1983).

















\end{thebibliography}
\end{document}